\documentclass[letterpaper,english,reprint,superscriptaddress,longbibliography]{revtex4-1}
\usepackage{lmodern}

\usepackage[T1]{fontenc}
\usepackage[utf8]{inputenc}
\usepackage{babel}
\usepackage{float}
\usepackage{textcomp}
\usepackage{amsmath}
\usepackage{amssymb}
\usepackage{graphicx}
\usepackage[unicode=true,pdfusetitle,
 bookmarks=false,
 breaklinks=false,pdfborder={0 0 0},pdfborderstyle={},backref=false,colorlinks=false]
 {hyperref}

\makeatletter

\pdfpageheight\paperheight
\pdfpagewidth\paperwidth

\providecommand{\tabularnewline}{\\}

\makeatother

\begin{document}

\title{A fiber coupled cavity QED source of identical single photons}

\author{H. Snijders}

\affiliation{Huygens-Kamerlingh Onnes Laboratory, Leiden University, P.O. Box
9504, 2300 RA Leiden, The Netherlands}

\author{J. A. Frey}

\affiliation{Department of Physics, University of California, Santa Barbara, California
93106, USA}

\author{J. Norman}

\affiliation{Department of Electrical \& Computer Engineering, University of California,
Santa Barbara, California 93106, USA}

\author{V. P. Post}

\affiliation{Huygens-Kamerlingh Onnes Laboratory, Leiden University, P.O. Box
9504, 2300 RA Leiden, The Netherlands}

\author{A. C. Gossard}

\affiliation{Department of Electrical \& Computer Engineering, University of California,
Santa Barbara, California 93106, USA}

\author{J. E. Bowers}

\affiliation{Department of Electrical \& Computer Engineering, University of California,
Santa Barbara, California 93106, USA}

\author{M. P. van Exter}

\affiliation{Huygens-Kamerlingh Onnes Laboratory, Leiden University, P.O. Box
9504, 2300 RA Leiden, The Netherlands}

\author{W. Löffler}
\email{loeffler@physics.leidenuniv.nl}

\affiliation{Huygens-Kamerlingh Onnes Laboratory, Leiden University, P.O. Box
9504, 2300 RA Leiden, The Netherlands}

\author{D. Bouwmeester}

\affiliation{Huygens-Kamerlingh Onnes Laboratory, Leiden University, P.O. Box
9504, 2300 RA Leiden, The Netherlands}

\affiliation{Department of Physics, University of California, Santa Barbara, California
93106, USA}
\begin{abstract}
A high-fidelity source of identical single photons is essential for
numerous quantum technologies such as quantum repeaters and optical
quantum information processing \cite{Santori2002,Yamamoto1999}. Hallmarks
thereof are a near-unity single-photon purity, near-unity indistinguishability
of consecutively emitted photons, and high brightness through a near-unity
number of photons per time bin \cite{Somaschi2015a,Ding,Aharonovich2016}.
In order to embed such sources in quantum networks, optical fiber
integration is essential but complicated by cryogenic compatability
and noise. Here we demonstrate a resonantly pumped, quantum dot (QD)
based, transmission operated, single-mode fiber coupled single photon
source with a purity of 97$\%$, indistinguishability of 90$\%$,
and a brightness of 17$\%$. This is achieved by deploying a unique
micropillar cavity design in a closed-cycle cryostat, which is operated
using a through-fiber cross-polarization technique to remove the pump
laser light from the resonantly scattered single photons. These results
pave the way for fully fiber integrated photonic quantum networks,
as our technology is equally applicable for cavity-QED based photonic
quantum gates \cite{Kim2013,Reiserer2014a}.
\end{abstract}
\maketitle
\textbf{}

For a single photon source, high brightness and on-demand availability
is crucial for efficient implementation of quantum photonic protocols.
Additionally, to exploit the power of quantum interference such as
in boson sampling, consecutively produced photons need to be indistinguishable,
meaning that their wave functions must overlap well. Until recently,
heralded spontaneous parametric down conversion sources \cite{Barz2010}
were state of the art for single photon sources (SPS) \cite{Eisaman2011},
with which most quantum communication and optical quantum computing
protocols have been demonstrated \cite{Takemoto2015}. The main problem
of these sources is that the Poissonian statistics of the generated
twin photons will always result in a trade-off between single-photon
purity (the absence of N>1 photon number states) and brightness (the
probability to obtain a photon per time slot).

One way to deterministically produce single photons is to use trapped
atoms \cite{Kuhn2002}, where single photon rates up to 200 kHz have
been obtained recently \cite{Higginbottom2016}. In order to enable
integration and increasing the photon rate, solid-state systems have
been investigated. An excellent candidate is a semiconductor QD \cite{Santori2002,Bennett2016a,Hennessy2007}.
QDs have nanosecond lifetime transitions that enable GHz rate production
of single photons as required for numerous quantum technologies. Compared
to other solid state emitters such as NV centers, nanowire QDs, excitons
in carbon nanotubes or two-dimensional materials \cite{Rogers2014,Sipahigil2014},
self-assembled QDs in cavities can show almost perfect purity and
indistinguishability \cite{Somaschi2015a}. A challenging task is
to couple the quantum emitter to propagating light fields with near-unity
efficiency. This can be achieved by placing them in optical micro
cavities, which additionally increases the emission rate by cavity-QED
Purcell enhancement, such as micropillar cavities \cite{Snijders2016,Santori2002},
photonic crystal cavities \cite{Muller2014}, or ring resonators
\cite{Davanco2016}.

For the next major step in implementing single photon sources in complex
photonic quantum networks, such as for quantum boson sampling or cluster
state quantum computation, coupling to a single mode optical fiber
is essential. Several challenges are connected to this: cryogenic
compatibility \cite{Haupt2010}, resonant optical pumping (a preprint
about a nonresonantly pumped device appeared recently \cite{Schlehahn2017}),
high coupling efficiency and robust and stable polarization control.
Another approach is to employ fiber-tip micro cavities but the photon
collection efficiency is limited to about 10\% to date \cite{Muller2009,Greuter2015a}.

\begin{figure}[h]
\includegraphics[scale=0.18]{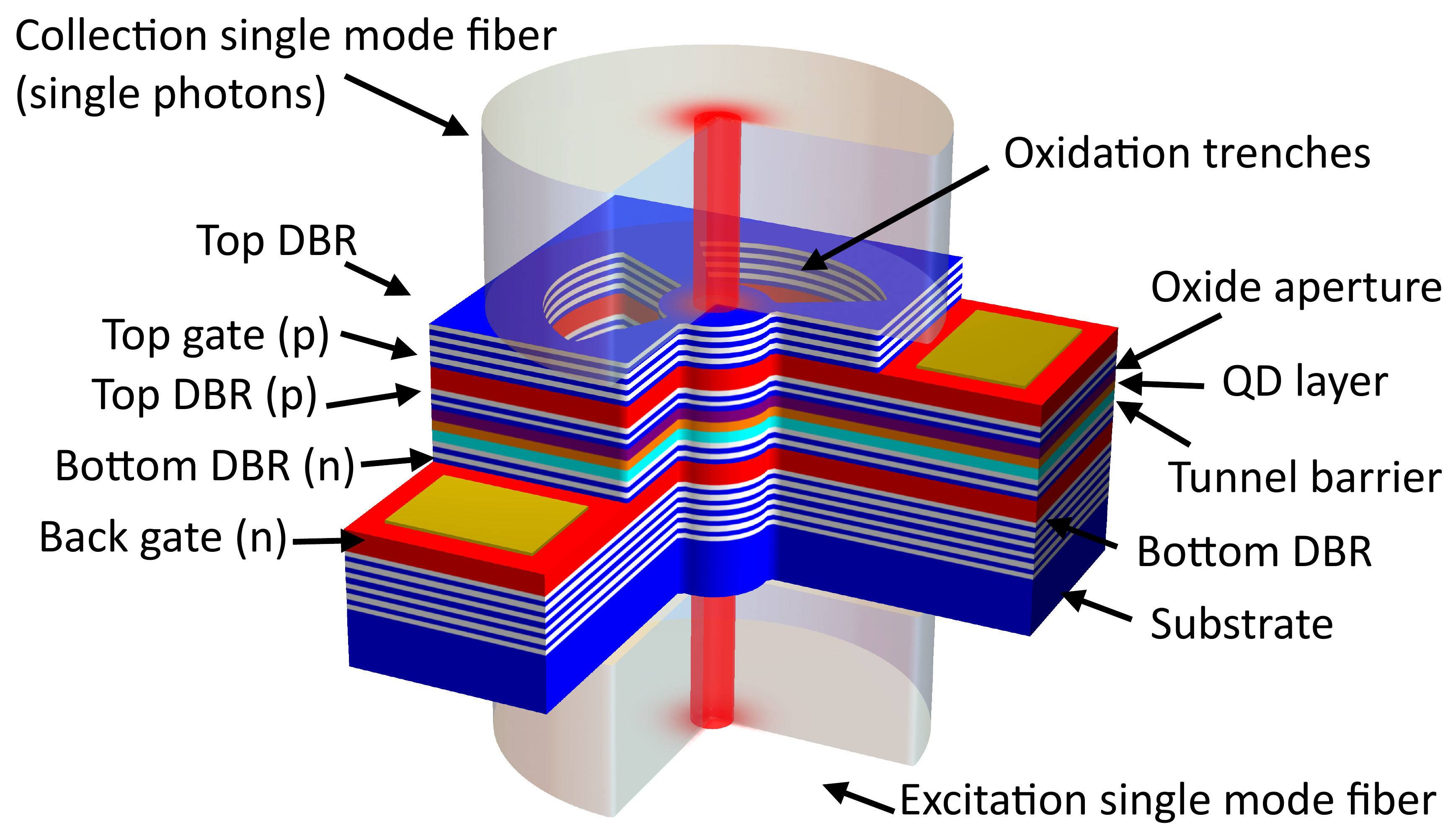}

\caption{Sketch of the micro-cavity quantum dot device with attached fibers
from bottom (excitation fiber) and top (single photon collection fiber).
The trenches are used for wet-chemical oxidation of a sacrificial
AlAs layer to form an intra-cavity lens or aperture that leads to
transverse confinement of the optical cavity mode. \label{FC-SPS-device}}
\end{figure}

\textbf{}Here, we show a prototype of a fully fiber coupled solid-state
resonantly pumped and transmission-based source of identical photons.
Our fiber coupled single photon device is sketched in Fig. \ref{FC-SPS-device}:
The device consists of a layer of self-assembled InAs/GaAs QDs embedded
in a micropillar Fabry–Perot cavity (Purcell factor $F_{p}=3.8$)
grown by molecular beam epitaxy \cite{Strauf2007a}. The QD layer
is embedded in a P–I–N junction, separated by a 27 nm thick tunnel
barrier from the electron reservoir to enable tuning of the QD resonance
frequency by the quantum-confined Stark effect. Since we do not use
air-guided micropillars but an oxide aperture for 3D confinement \cite{Bonato2012},
the device is very robust and the optical properties do not degrade
by attachment of the fibers. The singlemode fibers are carefully aligned
and connected by an UV-light curable adhesive to the front and back
of the device (Supplementary Section 1), allowing for transmission
measurements with full control of the optical polarization.

\begin{figure}
\includegraphics[scale=0.12]{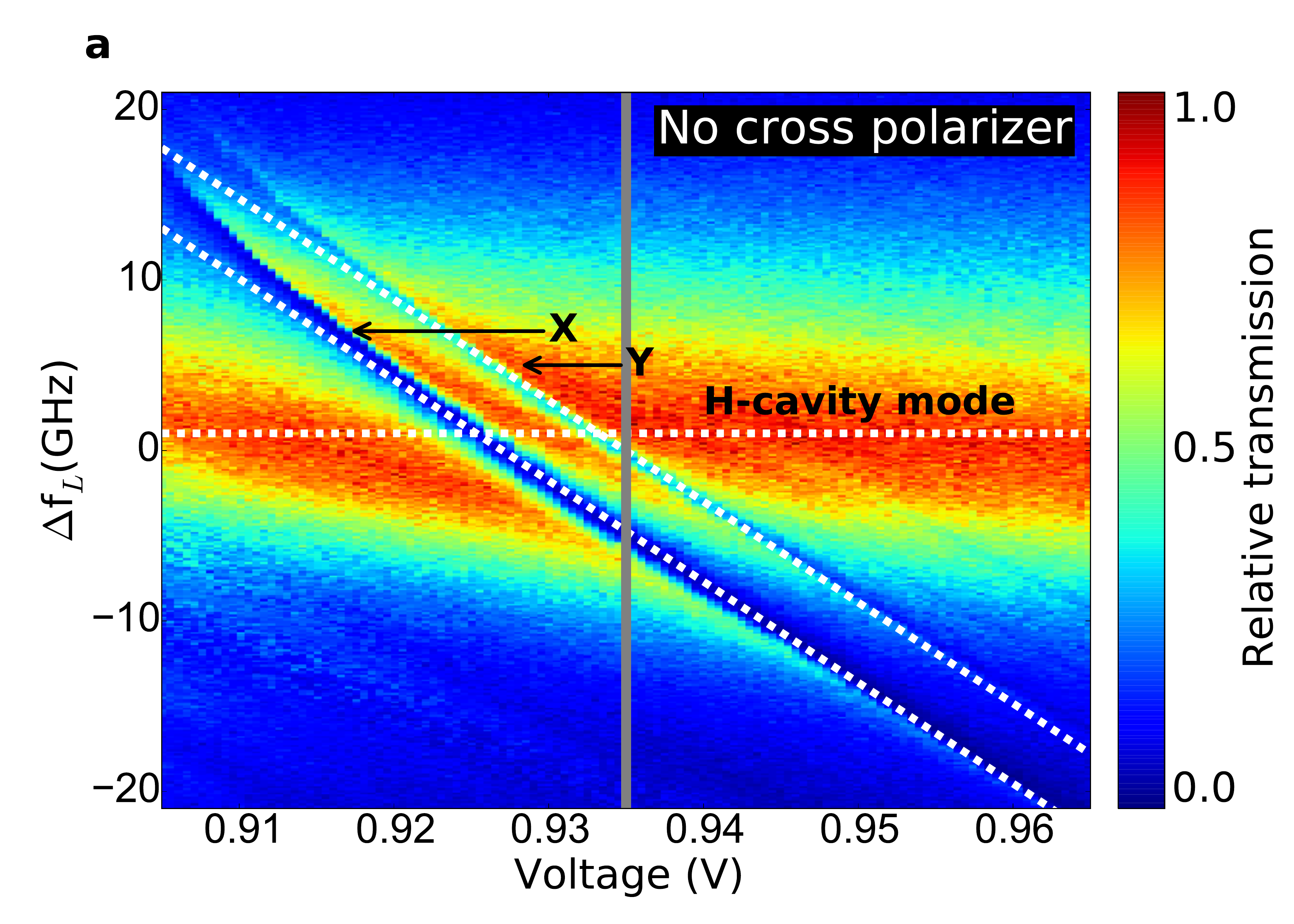}\includegraphics[scale=0.12]{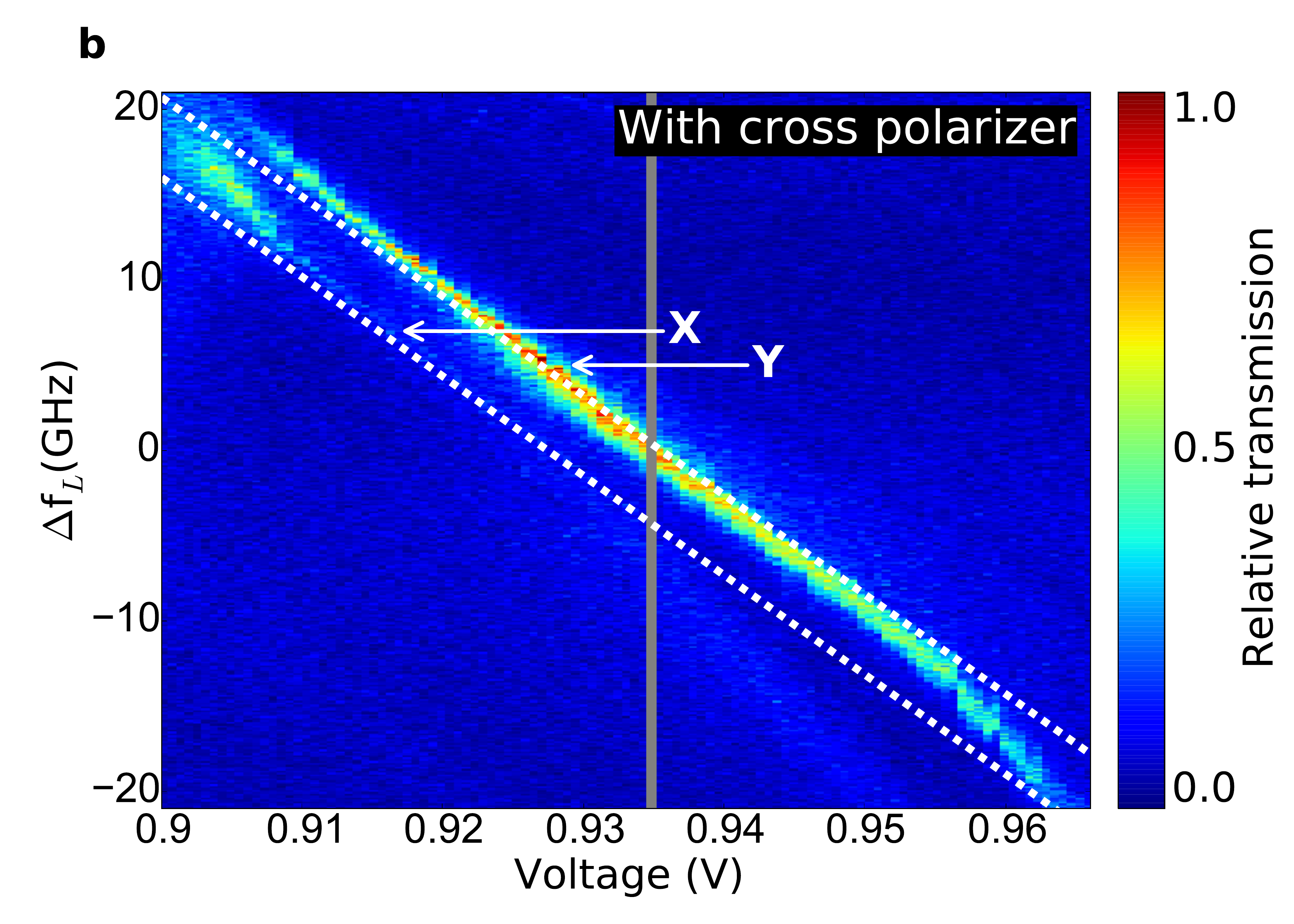}

\includegraphics[scale=0.15]{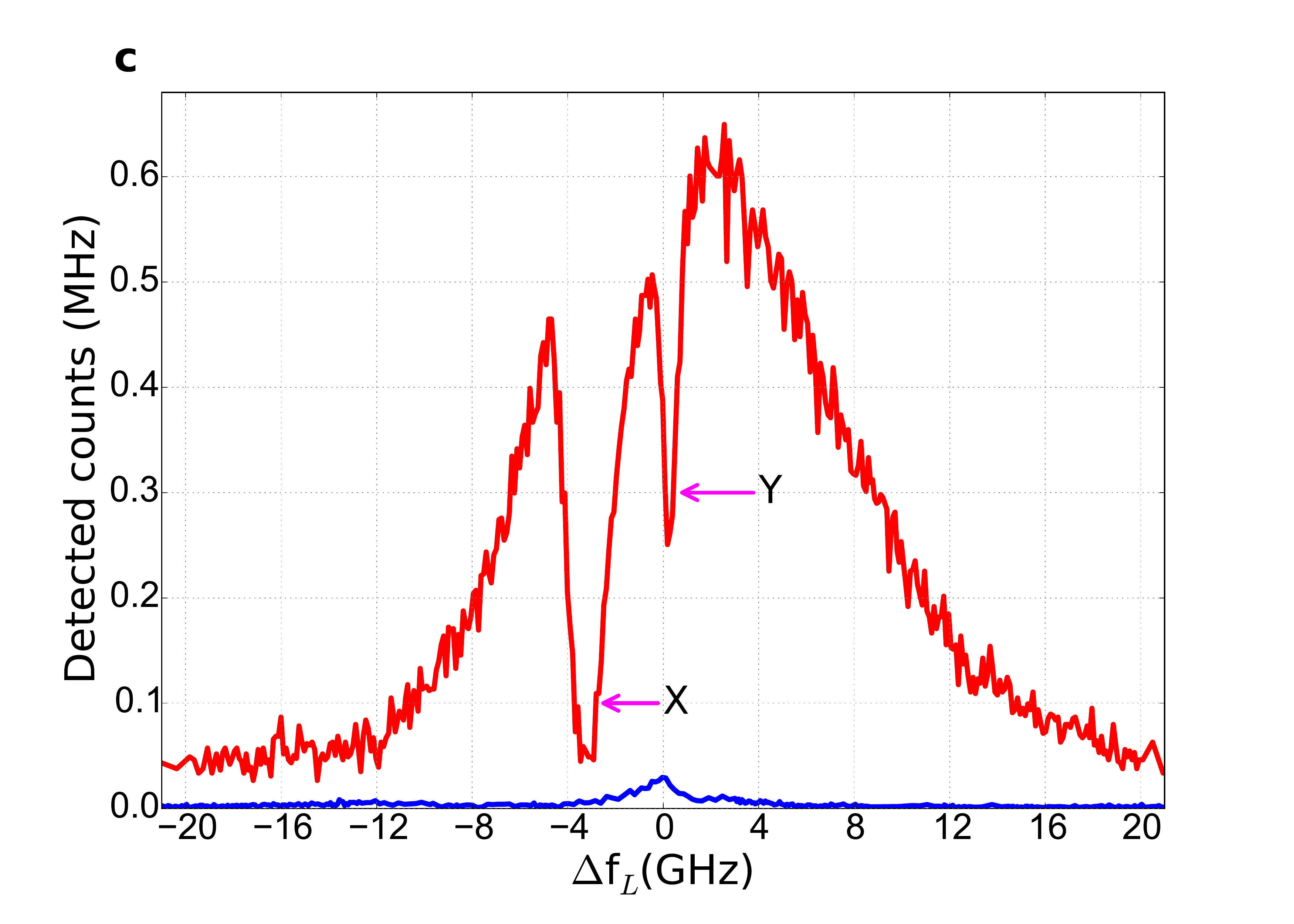}

\caption{a, b: False color plots of resonant transmission as a function of
laser frequency and gate voltage. In panel a, the incident laser light
is polarized along the H cavity axis, and the transmitted light is
detected without polarization selection. In b, the remnant laser light
is filtered out using a crossed polarizer oriented along the V-polarized
cavity mode, to select the photons coherently scattered from the Y-transition
of the QD. Panel\textbf{ c} shows cross sectional plots (red line:
without polarization selection, blue line: with crossed polarizer,
scan time $1\,$s) at a gate voltage of 0.935 V (grey line in a and
b). Indicated are the X and Y QD transitions and the H-polarized cavity
mode. \label{QD-measurements}}
\end{figure}

\textbf{}The fundamental cavity mode is split in two linearly polarized
modes, the H and V mode, induced by a small ellipticity of the cavity
cross-section. Similarly, the neutral exciton transition of the QD
is split in two linearly polarized transitions by the fine structure
exchange interaction. Fig. \ref{QD-measurements}a shows a false color
plot of the transmission through the sample as a function of the applied
voltage and laser frequency. Using a free-space polarizer and a fiber
polarization controller, the input polarization is set along the H
cavity polarization axes. The transmitted light is sent to a single
photon detector. The two fine structure split QD transitions are clearly
visible as dips in the transmission spectrum that shift as a function
of the applied electric field. A cross sectional plot of Fig. \ref{QD-measurements}a
(grey line)\textbf{ }is shown in Fig. \ref{QD-measurements}c (red
line). The depth of the dips indicate the ``X'' QD transition couples
better to the H cavity mode than the ``Y'' QD transition. This is
confirmed by comparison to a numerical model taking all relevant cavity-QED
and polarization effects into account (Supplementary Section 3). From
this model we also determine the angle $\theta$ between the X QD
axis and the H cavity mode axis to be $\theta=17$$^{\circ}$, and
the polarization splitting of the fundamental cavity mode (18 GHz). 

Fig. \ref{QD-measurements}b and c\textbf{ }(blue line)\textbf{ }show
single photons that are filtered from the transmitted light with a
combination of a fiber polarization controller and a free-space optical
polarizer set to extinguish the transmitted laser light (cross polarization).
We excite the system along the H cavity mode polarization but detect
only photons emitted from the V-polarized cavity mode. This is ideal
for efficient collection of the single photons that are coherently
scattered from the Y-transition of the QD, as is seen in Fig. \ref{QD-measurements}b.
This is a workable scheme because for excitation of the QD-cavity
system, we can simply remedy the reduced coupling of the Y QD transition
to the H-polarized cavity mode by increasing the laser power, while
the emitted single photons are very efficiently collected by the V-polarized
cavity mode. This also means that the Y QD transition acts here as
the better single photon source than the X QD transition. 

\textbf{}We now investigate the dependency between maximum single
photon rate and single photon purity that is achievable with the present
device. For this, we first perform continuous-wave resonant spectroscopy
experiments with a single frequency diode laser.
\begin{figure}[h]
\begin{raggedright}
\begin{minipage}[t]{0\columnwidth}%
\begin{center}
\includegraphics[scale=0.2]{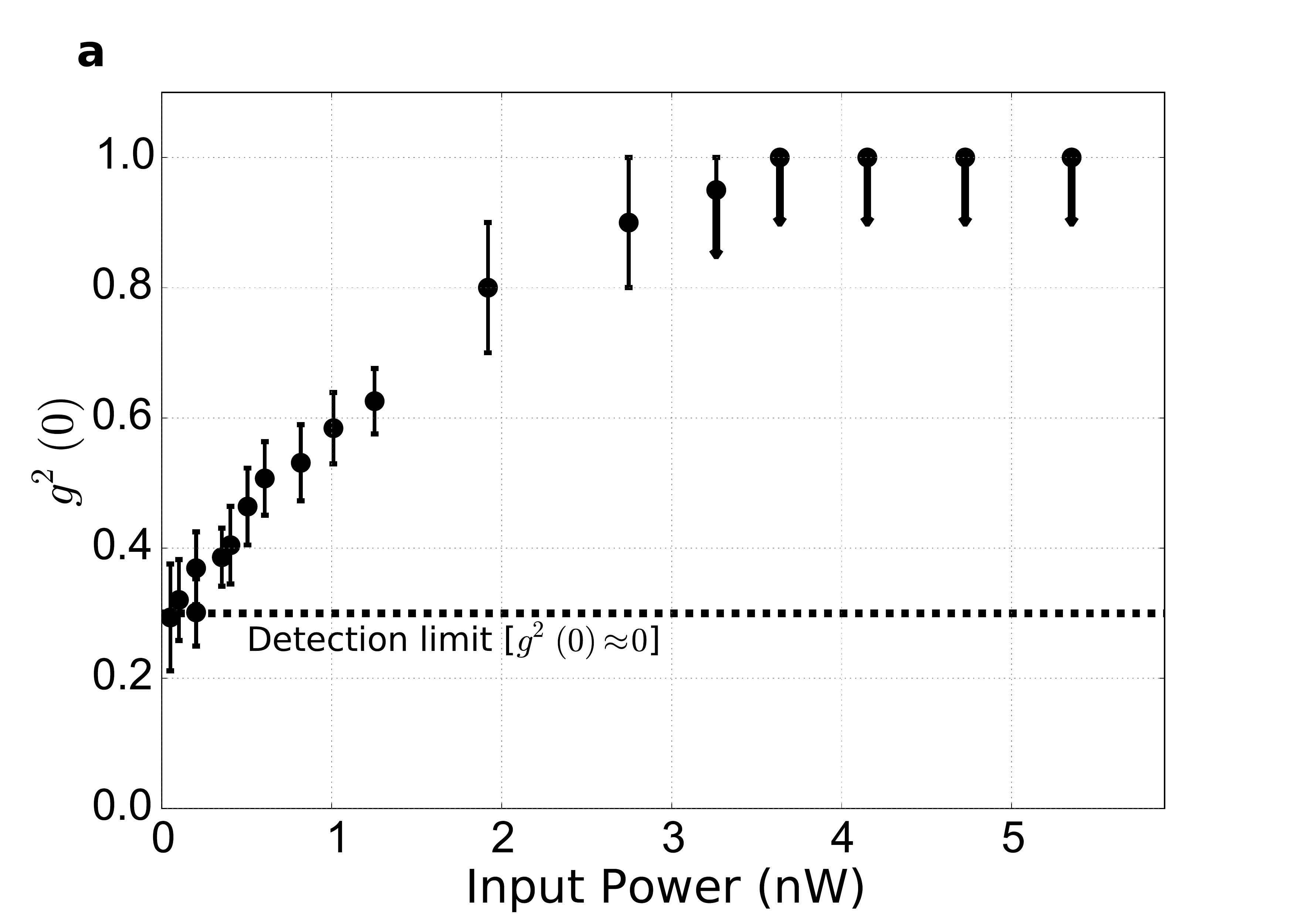}
\par\end{center}
\includegraphics[scale=0.2]{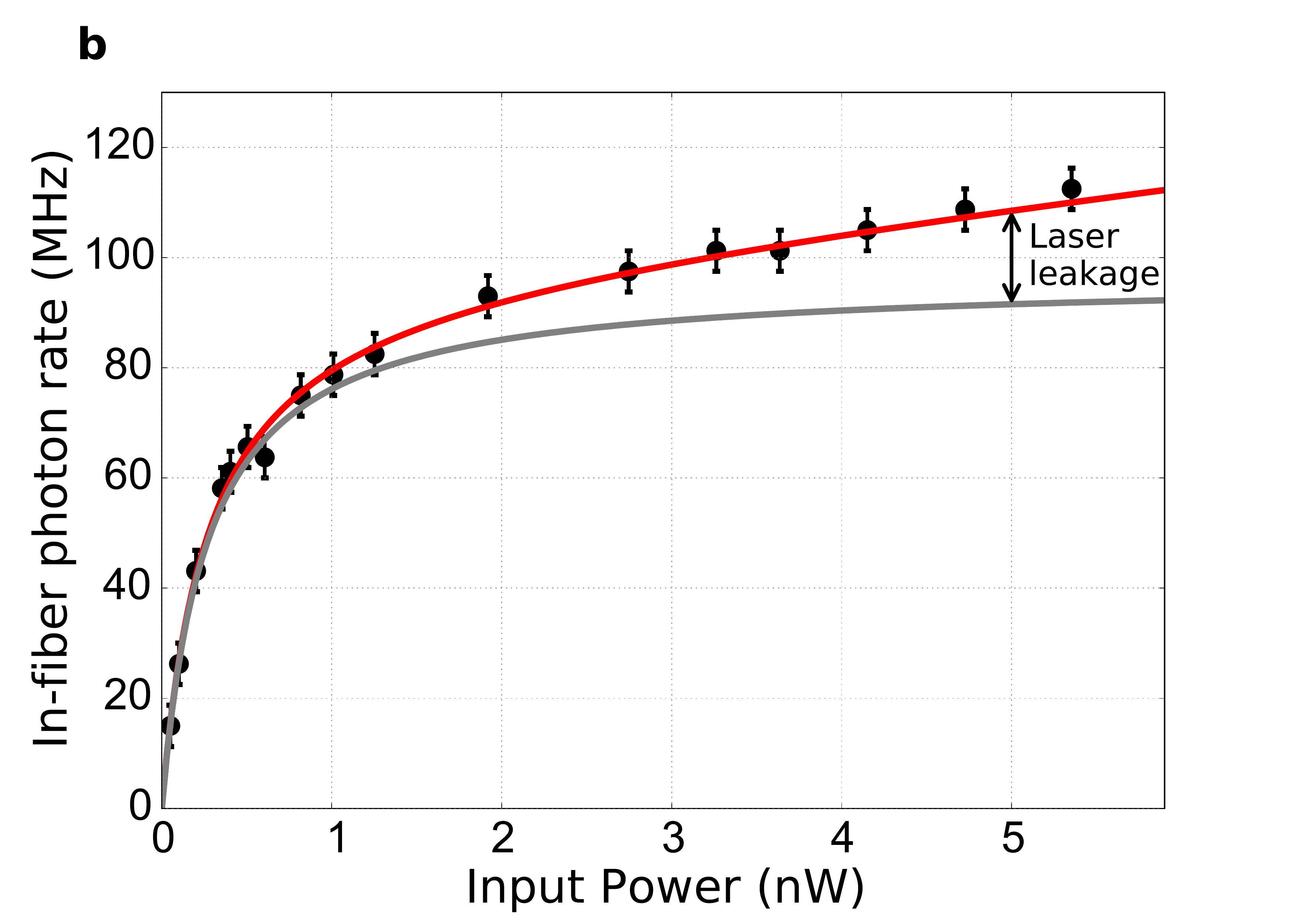}%
\end{minipage}
\par\end{raggedright}
\caption{a, Measurement of the second order correlation function $g^{2}(0)$
versus the incident laser power under continuous-wave excitation.
The dashed line indicates the limit on $g^{2}$ set by the detector
jitter, see Supplementary Section 4 for details. b, Simultaneously
measured single photon rate (corrected for detection efficiency).
The fit (red line) takes into account the saturation of the QD transition,
as well as residual laser light due to non-perfect polarization extinction.
\label{singlephotonrate}}
\end{figure}
We measure the second order correlation $g^{2}(0)$ and the flux of
emitted photons as a function of the incident laser power (Fig. \ref{singlephotonrate}a
\& b). This measurement is carried out under the above-mentioned cross-polarization
condition with an extinction ratio of about $10^{-4}$. Fig. \ref{singlephotonrate}a
shows the second order correlation function at $\Delta\tau=0$ as
a function of the incident laser power, while Fig. \ref{singlephotonrate}b
shows the emitted single photon rate measured simultaneously. For
increasing laser power, we observe a saturation of the single-photon
count rate at around 1 nW, which corresponds well to previous results
on related devices \cite{Bakker2015f}. Based on the bare lifetime
of the QD ($\gamma^{||}=1.0\pm0.4$ ns$^{-1}$) one would expect that
GHz rate single photons can be obtained, but we observe an increase
of $g^{2}(0)$ for increasing laser power. This is investigated in
Fig. \ref{singlephotonrate}b\textbf{ }(red line), where we have taken
into account the saturation response of the resonantly driven two
level system \cite{Lou}, and the laser leakage due to a non-perfect
optical polarization extinction. From this we can conclude that residual
coherent excitation laser light is responsible for the increase in
the second order correlation $g^{2}(0)$ at higher laser powers. 

\begin{figure}
\includegraphics{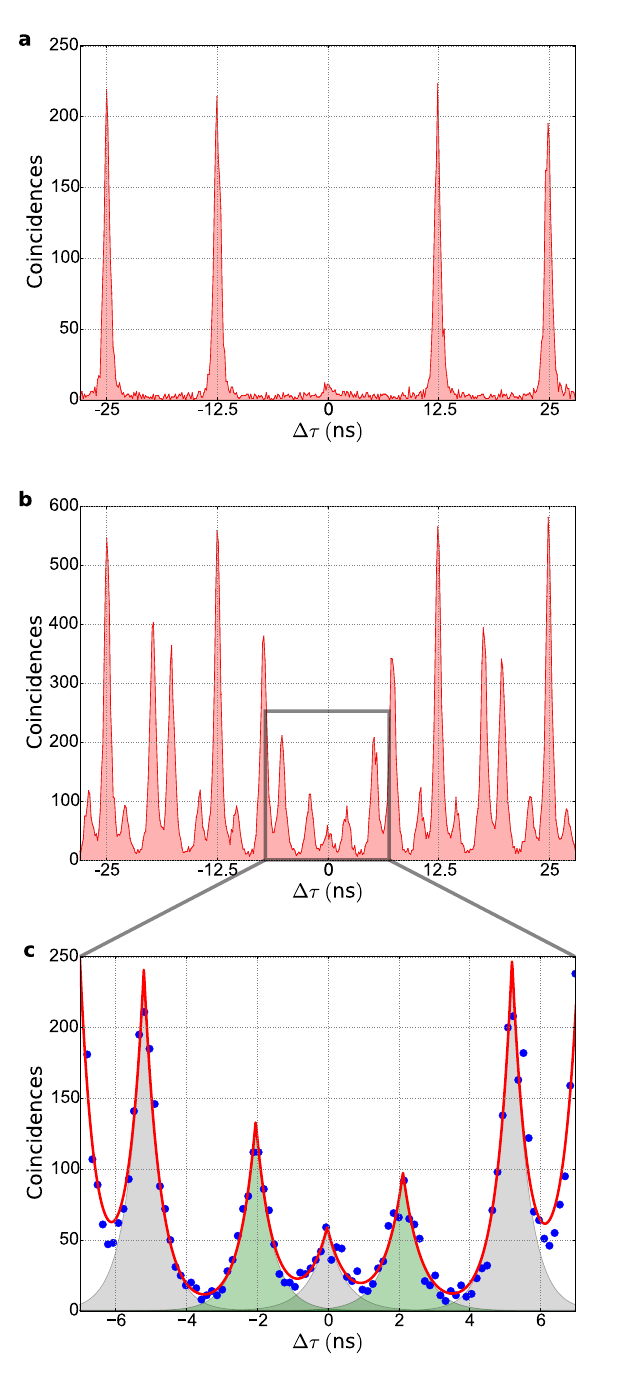}

\caption{\label{fig:Pulsed-laser}Photon correlations of the same QD transition
under pulsed excitation. In Fig. \ref{fig:Pulsed-laser}\textbf{a
}the second-order correlation $g^{2}(0)=0.037$ is calculated from
the integrated photon counts in the zero time delay peak divided by
the average of the adjacent four peaks. Fig. \ref{fig:Pulsed-laser}b
shows the indistinguishability measured for two pulses with a 5.2
ns delay. A magnified view around $\Delta\tau=0$ and a double exponential
fit of this data is given in Fig. \ref{fig:Pulsed-laser}c. Taking
into account a $g^{2}(0)=0.037$ we obtain a measured indistinguishability
of $M=0.90$. Measurement time: $600\,$s (a), $1200\,$s (b, c).}
\end{figure}

\textbf{}For quantum photonic applications, single photons are required
on demand with precise timing. We realize this using a resonant (around
932.58 nm) pulsed laser with 20 ps pulse length and 12.5 ns period.
These values are well-matched to the quantum dot transition in the
cavity as shown in Fig.~\ref{QD-measurements}c. Using a pulsed laser,
we are no longer limited by the jitter of the single photon detectors
and can obtain a more accurate value for $g^{2}(0)$. At a sufficient
low power of 100 pW, we measure a second order correlation of $g^{2}(0)=0.037\pm0.012$
as shown in Fig. \ref{fig:Pulsed-laser}a. Note that we did not use
spectral filtering of the cavity emitted light, in contrast to previous
investigations \cite{Somaschi2015a}. As we have investigated above,
$g^{2}(0)$ is in our case limited by imperfect extinction of the
excitation laser light. 

Next we determine the indistinguishability of two successively produced
single photons. For this, we send the emitted (single) photons into
a fiber-based Mach-Zehnder interferometer where one arm introduces
a delay of 5.2 ns. In order to create two excitation-laser pulses
with the exact same delay of 5.2 ns, we use a Michelson-type setup
with adjustable delay. As a result, consecutively emitted photons
arrive simultaneously at the final fiber splitter. We again measure
photon correlations between both output ports (Supplementary Section
2). If two consecutively produced single photons are indistinguishable,
they will undergo perfect quantum interference and ``bunch'', i.e.
two-photon coincidences at $\tau=0$ are expected to be absent in
the ideal case. This can be seen in Fig. \ref{fig:Pulsed-laser}b,
in particular if compared to the case where the photons are made distinguishable
artificially (Supplementary Section 5). By fitting the data with double
exponential functions and taking into account a finite value of $g^{2}(0)=0.037\pm0.012$,
we obtain an indistinguishability of $M=0.90\pm0.05$ (Fig. \ref{fig:Pulsed-laser}c).
Most of the coincidences that are still present at $\Delta\tau=0$
are due to the fact that the $g^{2}(0)$ is not completely zero. By
comparing the 80 MHz repetition rate of the laser to the detected
single photon rate, and by taking detection efficiency and loss carefully
into account, we obtain a brightness of the photons in the detection
fiber of $0.17\pm0.02$ photons per laser pulse. 

\textbf{}

\textbf{}

In conclusion, we have shown a prototype of a fully fiber coupled
solid-state single photon source that produces on-demand single photons
with a purity of $0.963\pm0.012$, indistinguishability of $0.90\pm0.05$
and a brightness of $0.17\pm0.02$. These figures are already very
promising for using such a device for quantum boson sampling or for
quantum-light spectroscopy applications \cite{Dorfman2016}. Furthermore,
we have demonstrated a first all-fiber coupled cavity-QED based photonic
quantum gate that filters out single photons from pulses of coherent
laser light with a scheme that is also compatible with more complex
excitation schemes \cite{Dada2016}. A next step is charging of the
QD with a single electron or hole spin to create a quantum memory
\cite{Kroutvar2004} which makes the device usable as a quantum node
for remote entanglement generation, quantum key distribution, and
distributed quantum computation.

\onecolumngrid%

\setcounter{table}{0}
\renewcommand{\thetable}{S\arabic{table}}
\setcounter{figure}{0}
\renewcommand{\thefigure}{S\arabic{figure}}%

\section*{Supplemental Information}

\section{Fiber coupled cavity QED device}

\begin{figure}[h]
\includegraphics{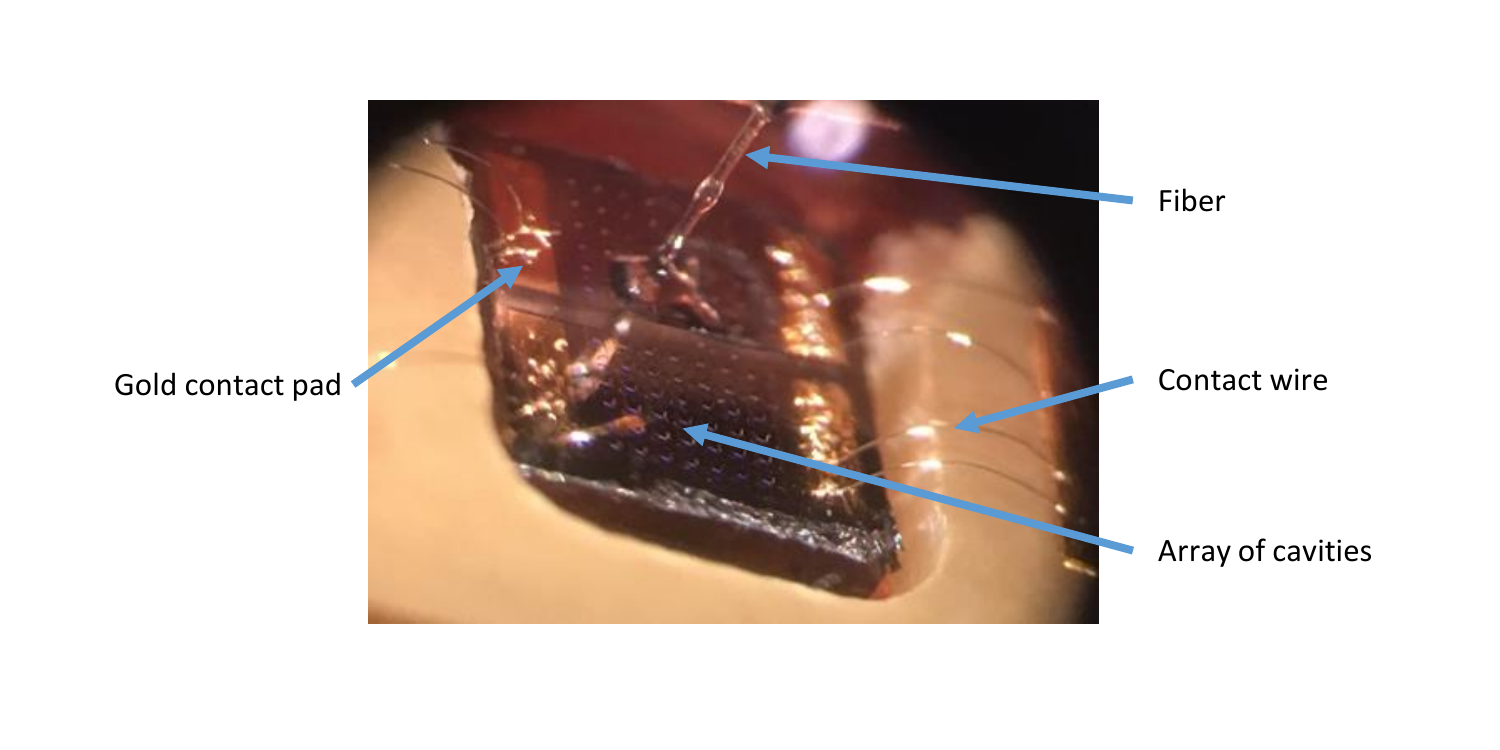}

\caption{\label{fig:cam-pics} Microscope image of the fiber coupled cavity-QED
device.}
\end{figure}

Fig. \ref{fig:cam-pics} shows a microscope image of the fiber coupled
cavity-QED device, visible is the front fiber attached to the sample
and the bond wires connected to the gold bond pads. 

\begin{figure}
\includegraphics[scale=0.5]{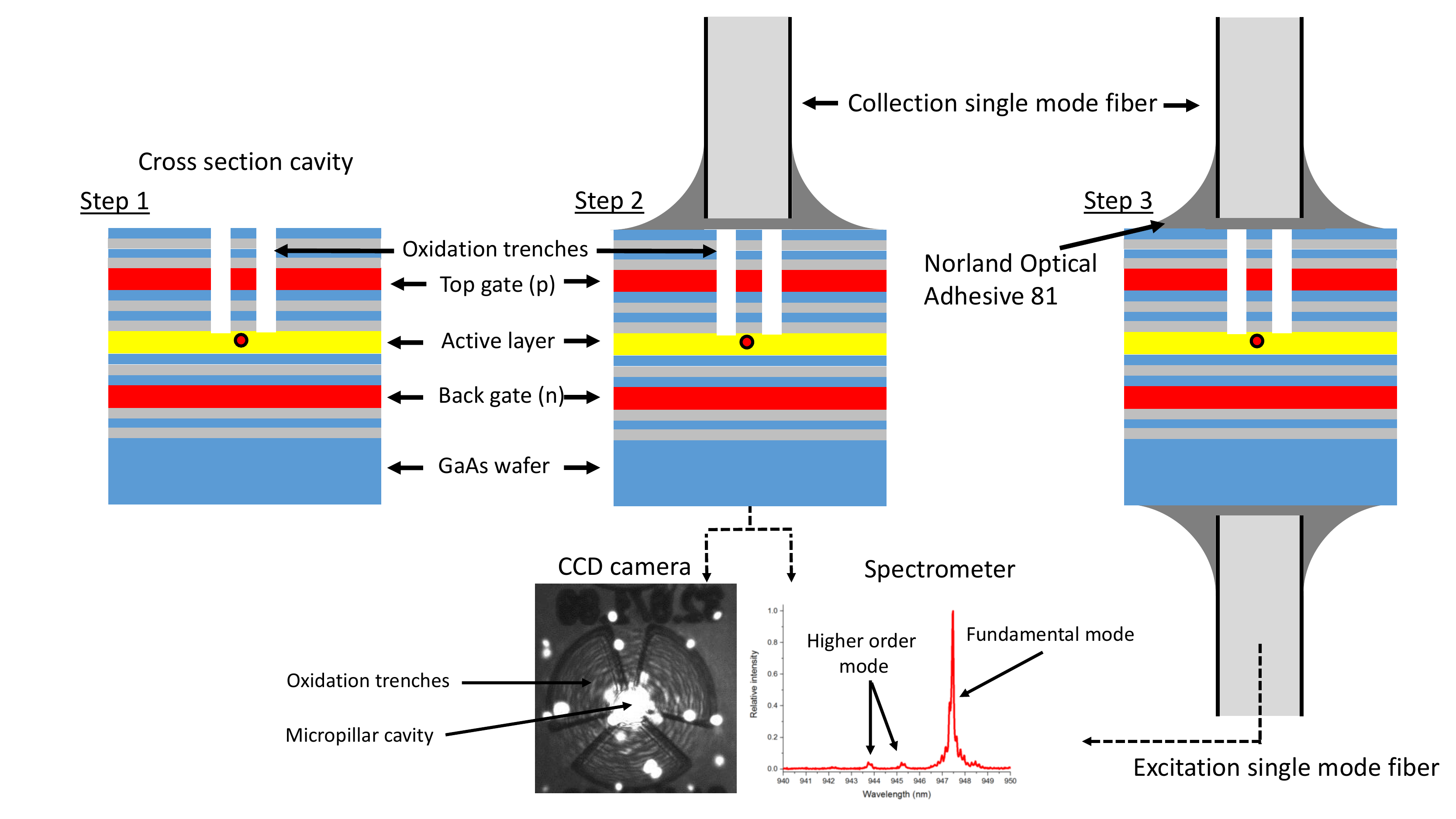}

\caption{\label{fig:attachfibers}Sketch of the procedure for connecting single
mode fibers to the cavity-QED device. }
\end{figure}

Attaching the fibers to the device requires three steps (see Fig.
\ref{fig:attachfibers}). 

\textbf{Step 1: }The sample is mounted in an optical spectroscopy
setup containing a long working distance microscope. The setup allows
for precisely aligning the single mode fibers with a motorized translation
stage. 

\textbf{Step 2:} Collection fiber attachment. Broadband light (900-980
nm) from a single mode fiber coupled superluminiscent diode is sent
into the collection fiber output. The other, cleaved, fiber end is
roughly positioned to the cavity by observing nonresonant transmitted
light using the microscope. For mode matching the fiber to the cavity
mode, the transmitted light is sent to a 1 m spectrometer. The optimal
position is found by maximizing the transmission in the fundamental
mode of the cavity and reducing the higher order TE/TM modes. After
finding the optimal position, the single mode fiber is vertically
moved up so that a drop of Norland Optical Adhesive 81 can be put
onto the cleaved fiber facet. After bringing the fiber back to its
original position, the adhesive should touch the sample which is verified
with an optical microscope. After again optimizing the position, the
adhesive is cured using UV-light. Before removing the device from
the setup, the fiber is firmly attached to the copper mount using
Stycast.

\textbf{Step 3:} Excitation fiber attachment. We flip the device around
and send again broadband light into the collection fiber. Rough alignment
of the excitation fiber at the bottom of the device is done by aligning
the fiber to the fundamental cavity mode using the microscope, after
which we use again a spectrometer to fine tune the position. Then,
the procedure from step 2 is used for attaching the fiber. 

The cavity mode of our device has at the front surface a waist of
$\omega_{front}=2.14\pm0.08\,$\textmu m and at the back a waist of
$\omega_{back}=28.48\pm1.02\,$\textmu m at around 955 nm \cite{Bonato2012}.
The increased waist at the back of the sample is due to the $650\,$\textmu m
thick GaAs wafer. The fibers (Thorlabs 780HP) have a core radius of
$2.2\,$\textmu m and 0.13 NA, which results in a mode waist of $\omega_{fiber}=2.95\pm0.25\,$\textmu m.
Neglecting the phase and only taking into account the mode waist of
the fiber, we have at the front side of the cavity a coupling efficiency
of \cite{GhatakK.Thyagarajan1998} 

\[
\eta=\left(\frac{2\omega_{fiber}\,\omega_{front}}{\omega_{fiber}^{2}+\omega_{front}^{2}}\right)^{2}\exp\left(-\frac{2u^{2}}{\omega_{fiber}^{2}+\omega_{front}^{2}}\right).
\]
Here, $u$ is the transverse misalignment distance. Setting $u=0$
we obtain an optimal efficiency of $\eta_{front}=90\%\pm7.6\%$. Experimentally,
we obtain for our device a coupling efficiency of $60\%$. The reason
for this deviation is most likely a slight misalignment of the fiber,
which happens during curing of the adhesive and cooling down of the
sample; this could be avoided by improved management of thermal expansion.
The fiber at the back of the sample has a reduced incoupling efficiency
of 0.6$\%$, due to the presence of the thick GaAs wafer. For operation
of our single photon source this reduced coupling efficiency is irrelevant
because we excite the system from the back where the coupling efficiency
only effects the required excitation laser power.

\section{Optical setup}

\begin{figure}[H]
\includegraphics[scale=0.5]{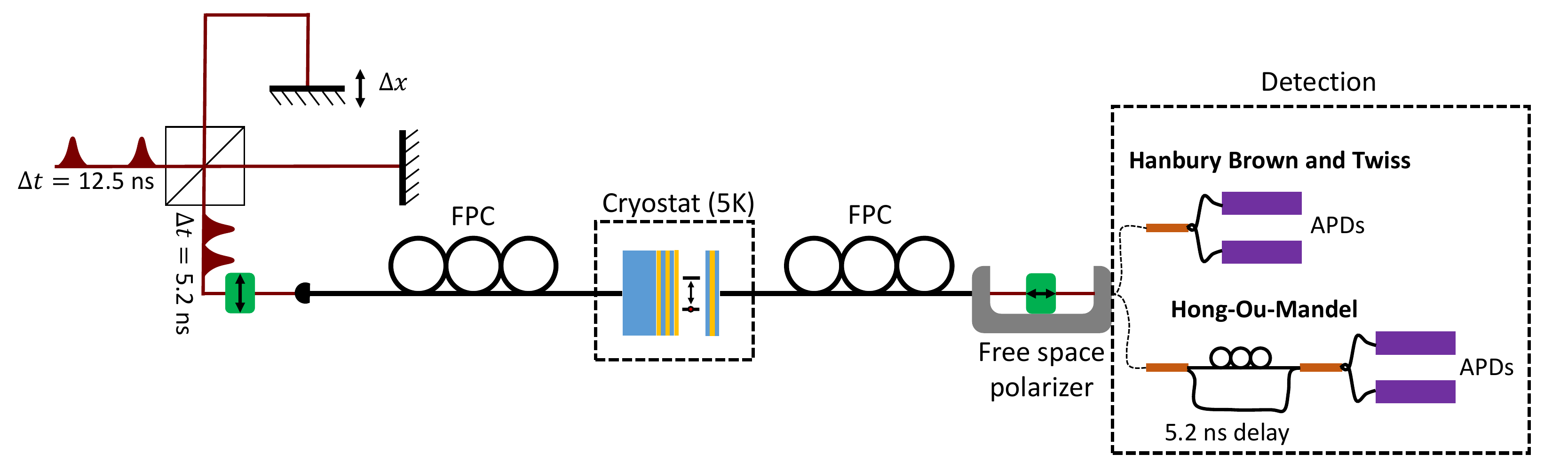}

\caption{\label{fig:sketchsetup}Sketch of the experimental setup. Dark red
lines indicate free space laser light at around 932.58 nm, single
mode fibers are depicted in black. FPC: fiber polarization controller.}
\end{figure}

The optical setup used to measure photon correlations to obtain single
photon purity and indistinguishability is shown in Fig. \ref{fig:sketchsetup}.
A pulse delay setup can be used to create from a mode-locked 80 MHz
Ti:Sa laser double pulses, which are sent to the micropillar cavity.
The transmitted photons are analyzed with a Hanbury Brown Twiss setup
to determine the second order correlation function $g^{2}(\Delta\tau)$,
or with a highly unbalanced Mach-Zehnder interferometer to observe
Hong-Ou-Mandel type photon bunching of consecutively emitted photons
to determine their indistinguishability $M$. Almost all components
in the setup are fiber-based or fiber-coupled, except the production
of the double laser pulses and the polarizers (fiber U-benches). The
delay between the double pulses is precisely adjusted to the Mach-Zehnder
interferometer delay by scanning $\Delta x$ while observing first-order
interference in absence of the cavity-QED device. This interference
signal becomes maximal when the position of $\Delta x$ matches the
in-fiber delay of about 5.2 ns.

\section{System Parameters}

In order to theoretically model the quantum dot cavity-QED system
we use an extended version of a model for a two level system in an
optical cavity driven by a classical coherent laser field. Using Qutip
\cite{Johansson2012a,Johansson2013} we solve numerically the quantum
master equation in the rotating wave approximation. Details about
the model we use to fit the data can be found in Ref. \cite{Snijders2016}. 

\begin{figure}[h]
\includegraphics[scale=0.2]{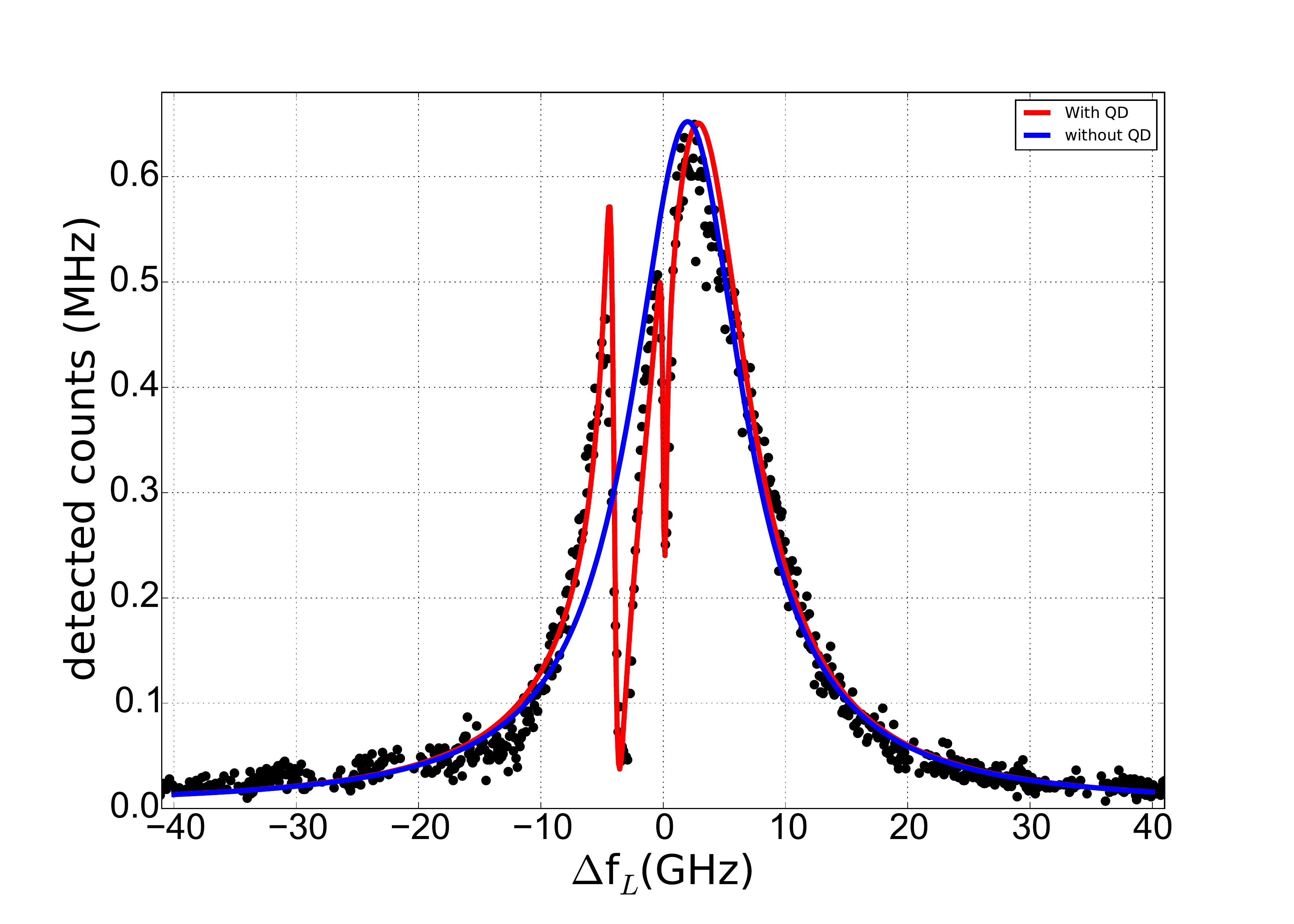}

\caption{\label{fig:fitssystem}Resonant transmission data of the two fine
structure split quantum dot transitions in a polarization non-degenerate
cavity. Black dots: experimental data, red line: theoretical model,
blue line: empty cavity.}
\end{figure}

We iteratively fit the simulation results to experimental data (shown
in Fig. \ref{fig:fitssystem}).\textbf{ }We obtain a cavity splitting
of $f_{cavsplit}=18\pm0.5$ GHz and a cavity decay rate $\kappa=70\pm3$
ns$^{-1}$. Now we keep these parameters fixed and optimize the model
for the case when only the H-polarized cavity mode is excited to obtain
the remaining 4 parameters of our QD-cavity system (\ref{fig:fitssystem}).
We find a QD-cavity coupling constant $g=14\pm0.4$ ns$^{-1}$, a
population relaxation rate of $\gamma^{||}=1.0\pm0.4$ ns$^{-1}$,
a pure dephasing rate of $\gamma^{*}=0.4\pm0.3$ ns$^{-1}$, a QD
fine structure splitting of $f_{QDsplit}=3.9\pm0.05$ GHz, and for
the angle between the H-polarized cavity mode and the X QD transition
$\phi=17^{\circ}\pm2^{\circ}$. The frequencies of the two fine-structure-split
QD transitions in Fig. \ref{fig:fitssystem} are $f_{QDX}=-3.6$ GHz
and $f_{QDY}=0.3$ GHz, and the frequencies of the polarization split
fundamental cavity modes are $f_{CavH}=2.0$ GHz and $f_{CavV}=20.0$
GHz.

From these parameters we find a cooperativity of $C=\frac{g^{2}}{\kappa(\gamma^{||}/2+\gamma^{*})}\approx2.8$
which corresponds to a Purcell enhancement of the excited state decay
rate of $F_{p}=C+1\approx3.8,$ assuming that the QD transition is
on resonance with the cavity transition. 

\section{Detector response}

\begin{figure}[H]
\includegraphics[scale=0.2]{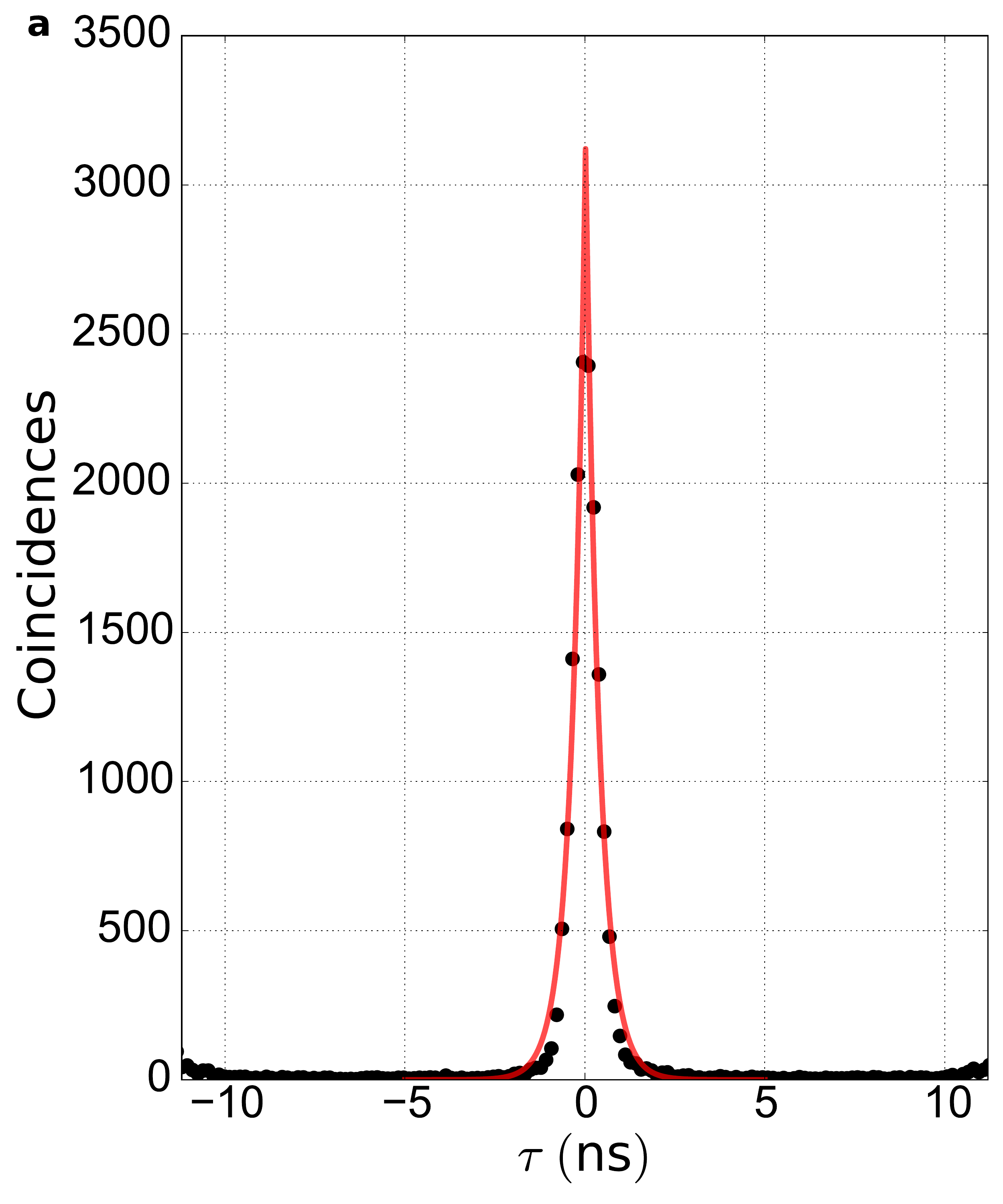}\includegraphics[scale=0.2]{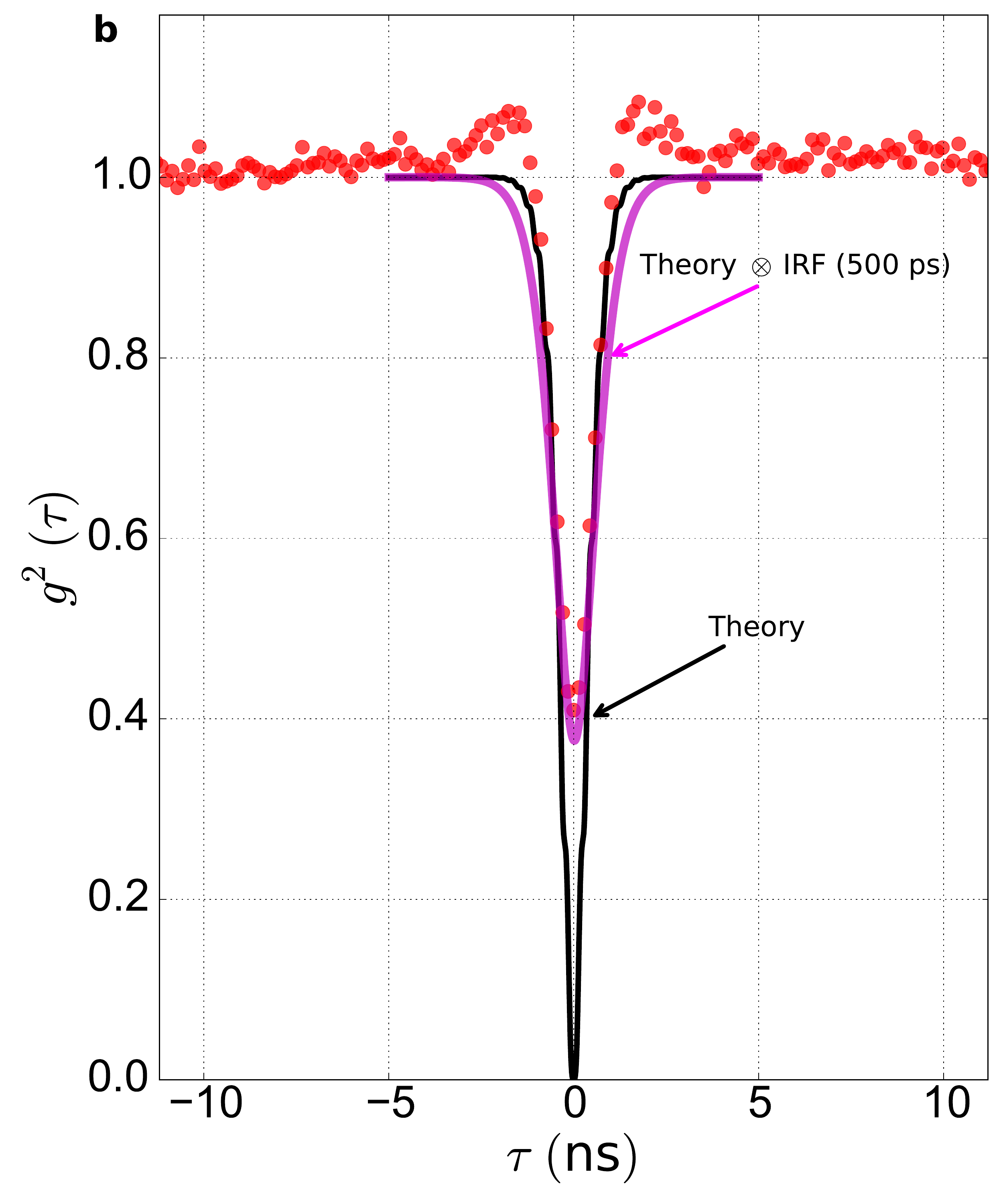}

\caption{\label{fig:detectresponse}\textbf{a }The two-detector response to
a very short light pulse is well fitted by a double-exponential function.\textbf{
b} Comparison of the experimental data with the convolution (purple
curve) of the detector response with the theoretical prediction (black
curve). }
\end{figure}

The two-detector response for our single photon counting detectors
(SPCM-AQR-14) is given by a double exponential function (Fig. \ref{fig:detectresponse}a).
Convoluting the theoretical prediction with this detector response
enables us to predict $g^{2}(0)$ for a continuous wave laser. As
is shown in Fig \ref{fig:detectresponse}b,\textbf{ }this agrees well
with the experimental data (red dots \ref{fig:detectresponse}b).
This proves that $g^{2}(\tau)$ measured with a continuous wave laser
is limited by detector jitter. 

\section{Indistinguishability measurement analysis}

\begin{figure}[h]
\includegraphics[scale=0.34]{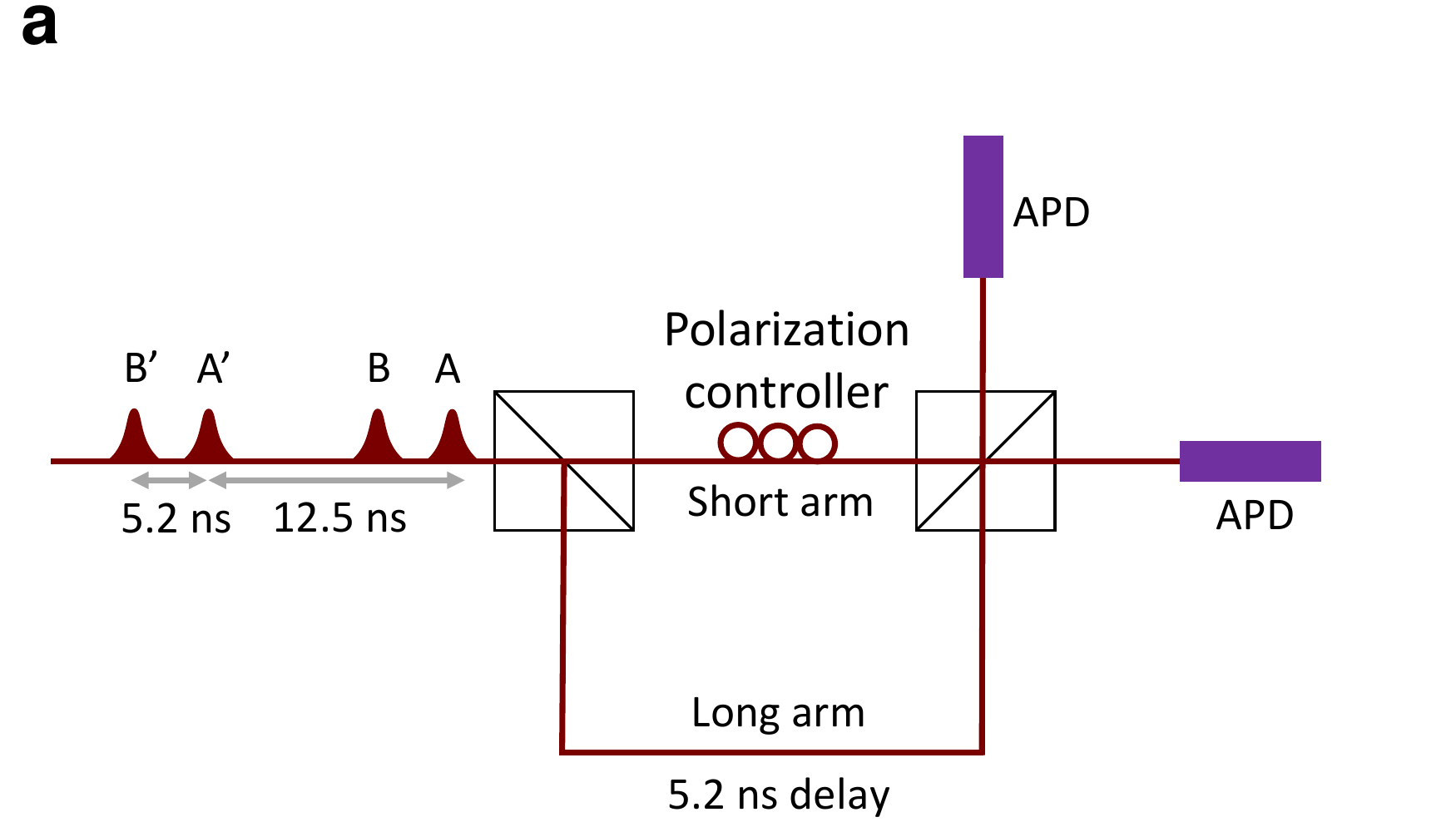}\includegraphics[scale=0.17]{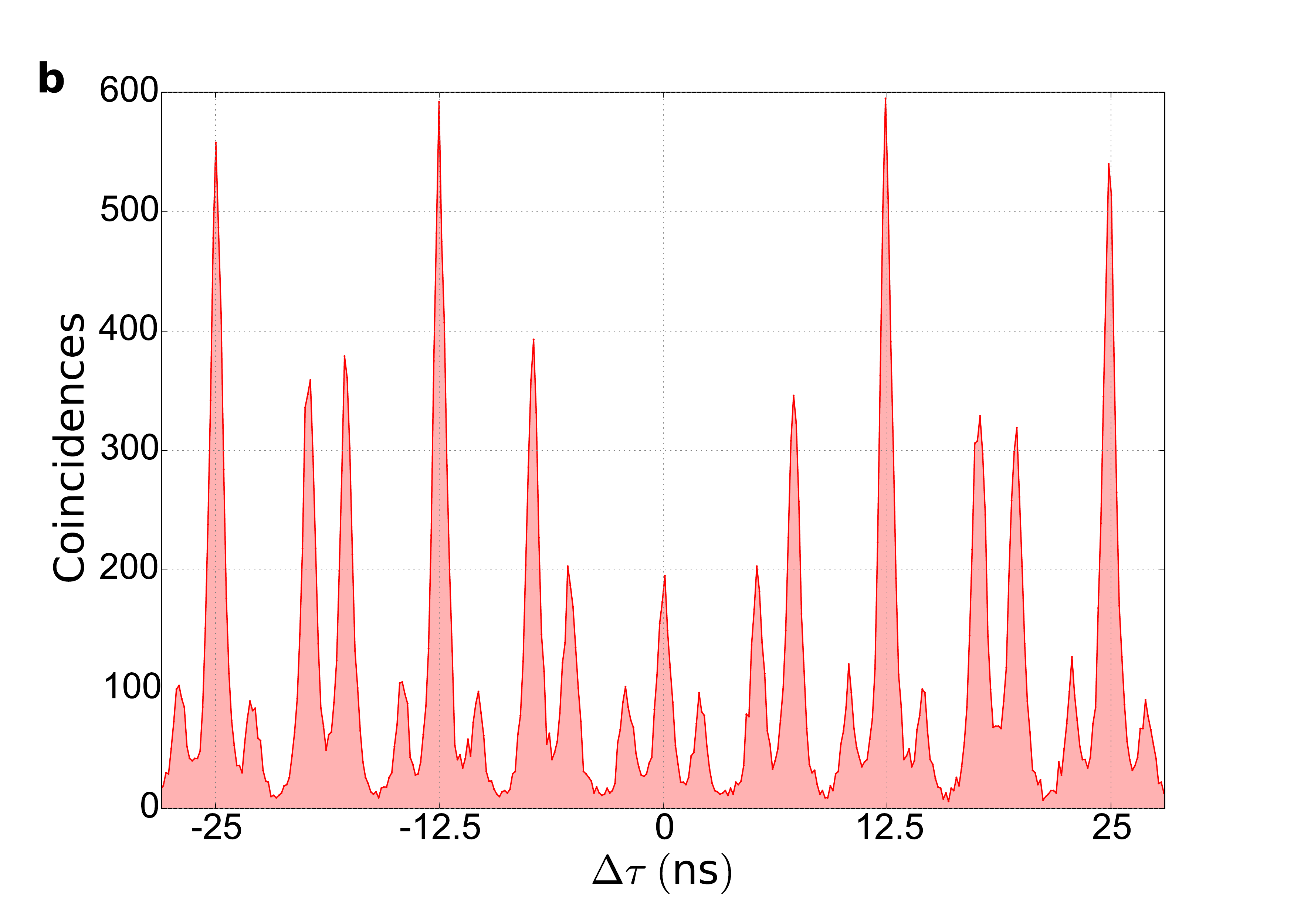}\includegraphics[scale=0.17]{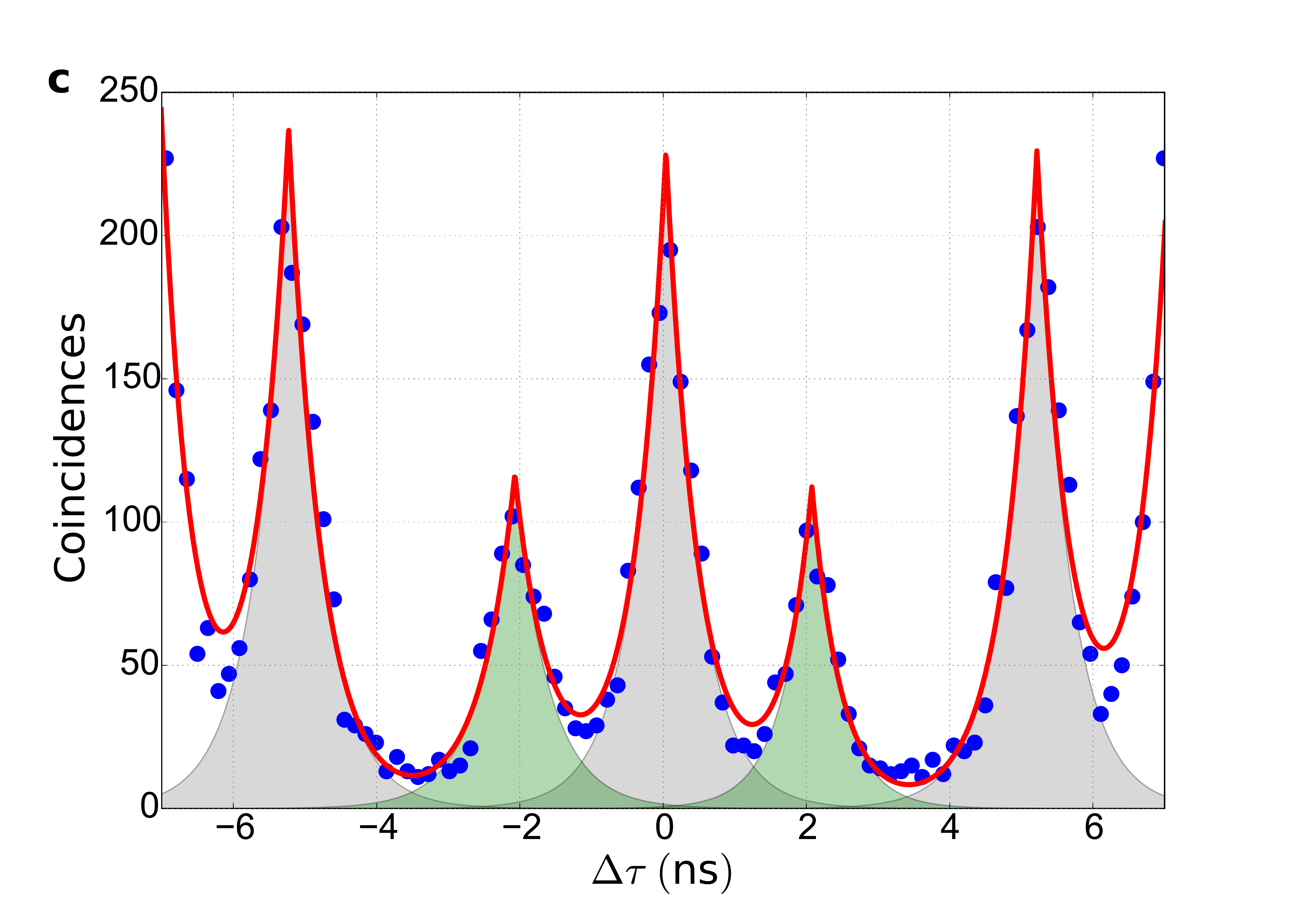}

\caption{\label{fig:Correlation-indist}\textbf{a} Detection scheme for measuring
the indistinguishability of consecutive photons. To compare to the
case of distinguishable photons, we rotate the polarization in one
arm of the interferometer. The result for distinguishable photons
is shown in (\textbf{b}, \textbf{c}), where photon bunching is suppressed
and two-photon coincidences at $\Delta\tau=0$ appear.}
\end{figure}

Here we explain the procedure to analyze the indistinguishability
measurements. First, we examine the ideal case without losses and
with ideal single-photon pulses (unity single-photon purity). We assume
an excitation laser pulse spacing of 5.2 ns. We need to consider two
double pulses and we label the photons as shown in Fig. \ref{fig:Correlation-indist}a:
photon A at 0 ns, B at 5.2 ns, A' at 12.5 ns, and photon B' at 17.2
ns. 

The detection is done using an unbalanced Mach Zehnder interferometer,
where one arm introduces a delay equal to the pulse delay (5.2 ns).
Photon correlations behind the last fiber splitter are measured. We
list all possible combinations of photons for which a two-photon detection
event can happen with a particular temporal delay between the photons.
In table \ref{tab:delays} below, the first row indicates the delay
between all possible photon combinations before the Mach-Zehnder interferometer.
The lower 4 rows show the four possible pathways which pairs of photons
can take, the number gives their relative delay at the single photon
detectors. The number of occurrences of a particular delay time is
directly proportional to the detection probability. For example, it
is 2 times more likely to detect two photons with $\Delta\tau=$5.2
ns than it is with 2.1 ns, which agrees very well with the experimental
data in Fig.~4c (main text). 

\begin{table}[H]
\vspace{0.5cm}

\begin{centering}
\begin{tabular}{|c|c|c|c|c|}
\hline 
 & AA' (ns) & BB' (ns) & BA' (ns) & AB (ns)\tabularnewline
\hline 
Laser pulse delay (before detection in MZ) & 12.5 & 12.5 & 7.3 & 5.2\tabularnewline
\hline 
\hline 
first photon long arm & 7.3 & 7.3 & 2.1 & 0\tabularnewline
\hline 
both photons short arm & 12.5 & 12.5 & 7.3 & 5.2\tabularnewline
\hline 
both photons long arm & 12.5 & 12.5 & 7.3 & 5.2\tabularnewline
\hline 
first photon short arm & 17.2 & 17.2 & 12.5 & 10.4\tabularnewline
\hline 
\end{tabular}
\par\end{centering}
\caption{\label{tab:delays} Table of arrival time differences $\Delta\tau$
of two-photon detection events. }
\end{table}

\vspace{0.5cm}

Indistinguishability: If the two photons are indistinguishable and
arrive simultaneously at the last fiber splitter, Hong-Ou-Mandel quantum
interference leads to photon bunching and prevents detection of coincidence
events, therefore, the ``AB'' event with $\Delta\tau=0$ in table
\ref{tab:delays} disappears. 

To contrast the indistinguishability measurement shown in Fig. 4c
in the main text to the case where the photons are perfectly distinguishable,
we perform an experiment where the photons are made artificially different
by giving them orthogonal polarization. The result in Fig. \ref{fig:Correlation-indist}b
clearly shows the absence of Hong-Ou-Mandel type photon bunching by
the strong correlations at $\Delta\tau=0$. Fig. \ref{fig:Correlation-indist}c
shows a zoom-in with double-exponential fits to the measured data.
This agrees excellently to the expectation in table \ref{tab:delays},
note that the $\Delta\tau=0$ probability should be multiplied with
2 due to the coincidence of $\pm\Delta\tau$.

This model can be improved by taking into account the losses of the
fiber splitters (we assume that both fiber splitters are identical)
and a finite purity of the single photon pulses. To do this, we follow
the procedure of Ref. \cite{Santori2002}: The probability for a detection
event at the center peak normalized by the repetition rate and detection
efficiency is given by 

\[
A_{CP}=(R^{3}T+TR^{3})(1+2g^{2}(0))-2(1-\varepsilon)^{2}MT^{2}R^{2},
\]
where $M$ is the mean wave function overlap or indistinguishability,
$(1-\varepsilon)$ is the visibility of the Mach-Zehnder interferometer
and $R$ and $T$ are the reflection and transmission coefficients
of the fiber splitters. Comparing this to the probability for a detection
event at $\Delta\tau=\pm5.2$ ns, we obtain

\[
M=\frac{1}{(1-\varepsilon)^{2}}\frac{R^{2}+T^{2}}{2RT}\left[(1+2g^{2}(0))-\frac{A_{CP}}{A_{-5.2ns}+A_{5.2ns}}(2+2g^{2}(0))\right].
\]
For our fiber splitters we find $R=0.469$, $T=0.531$, $(1-\varepsilon)=0.96\pm0.1$.
We determine the coincidence probability ratio obtained from double
exponential fits in Fig. 4c\textbf{ }(main text) is $\frac{A_{CP}}{A_{-5.2ns}+A_{5.2ns}}=0.12\pm0.004.$
Combined with the single-photon purity measurement with $g^{2}(0)=0.037\pm0.012$
(main text, Fig. 4a), we obtain for the indistinguishability $M=0.90\pm0.05.$

\begin{acknowledgments}
We thank D. Kok and M.F. Stolpe for fruitful discussions. We acknowledge
funding from FOM-NWO (08QIP6-2), from NWO/OCW as part of the Frontiers
of Nanoscience program, and from the National Science Foundation (NSF)
(0901886, 0960331).
\end{acknowledgments}

\section*{Author contributions}

JAF, JN, AG, JEB, DB, and WL designed and fabricated the devices;
HS, VP, MPE, DB, and WL conceived and conducted the optical experiments.
All authors contributed to the manuscript.

\newpage{}

\bibliographystyle{naturemagwV1allauthors}
\bibliography{Mybibliography1}

\end{document}